\documentclass[
reprint,
twocolumn,
bibnotes,
amsmath,amssymb,
]{revtex4-2}
\usepackage{subcaption}
\usepackage{tikz}
\usepackage[font=footnotesize]{caption}
\usepackage{subcaption} 
\usepackage{mwe}
\usepackage{graphicx}
\usepackage{dcolumn}
\usepackage{bm}
\usepackage{hyperref}

\DeclareRobustCommand{\orcidicon}{%
	\begin{tikzpicture}
		\draw[lime, fill=lime] (0,0)
		circle [radius=0.16]
		node[white] {{\fontfamily{qag}\selectfont \tiny ID}};
		\draw[white, fill=white] (-0.0625,0.095)
		circle [radius=0.007];
	\end{tikzpicture}
	\hspace{-2mm}
}

\foreach \x in {A, ..., Z}{%
	\expandafter\xdef\csname orcid\x\endcsname{\noexpand\href{https://orcid.org/\csname orcidauthor\x\endcsname}{\noexpand\orcidicon}}
}

\DeclareRobustCommand{\orcidicon}{%
	\begin{tikzpicture}
		\draw[lime, fill=lime] (0,0)
		circle [radius=0.16]
		node[white] {{\fontfamily{qag}\selectfont \tiny ID}};
		\draw[white, fill=white] (-0.0625,0.095)
		circle [radius=0.007];
	\end{tikzpicture}
	\hspace{-2mm}
}

\foreach \x in {A, ..., Z}{%
	\expandafter\xdef\csname orcid\x\endcsname{\noexpand\href{https://orcid.org/\csname orcidauthor\x\endcsname}{\noexpand\orcidicon}}
}

\begin{document}
	
	
	\title{Photoinduced topological phase transition in monolayer 1T$^\prime$-MoS$_2$}

	\author{Mohammad Mortezaei Nobahari\orcidA}
	\email{mortezaie.mm71@gmail.com}
	\author{Mahmood Rezaei Roknabadi\orcidB}
	\email{roknabad@um.ac.ir}
	
	\affiliation{
		Faculty of science, Department of Physics, Ferdowsi University of Mashhad, Iran}

	
	
	
	\date{\today}
	
	\begin{abstract}

We investigate the systematic competition between perpendicular electric field and circularly polarized light in nonequilibrium topological phases of monolayer 1T$^\prime$--MoS$_2$ using a low-energy $k\!\cdot\!p$ Hamiltonian and Floquet theory. The off-resonant field generates spin- and valley-dependent mass corrections that reshape the Berry curvature (BC) and shift the conditions for band inversion. By evaluating the quasienergy bands, BC, and spin- valley-resolved Chern numbers, we identify a sequence of light and electric field controlled topological transitions by gap closing. Depending on the Floquet coupling strength and the electric-field parameter, the system evolves between the equilibrium quantum spin Hall (QSH) state and a set of driven phases including spin-polarized quantum Hall insulator (S-QHI), band insulator (BI) and photo-induced quantum Hall insulator (P-QHI) regimes. The results establish 1T$^\prime$--MoS$_2$ as a tunable platform where circular driving selectively manipulates spin and valley degrees of freedom, enabling controlled access to non-equilibrium topological phases in transition metal dichalcogenides (TMDs).

	\end{abstract}
	
	\maketitle
	
	
	\section{\label{sec:1}Introduction }

                       The discovery of topological phases of matter has profoundly reshaped modern condensed matter physics, uncovering quantum states that are characterized not merely by broken symmetries but by global topological invariants defined in momentum space. Among these, the quantum spin Hall (QSH) effect stands as a cornerstone of two-dimensional (2D) topological physics, representing a time-reversal-symmetric analogue of the quantum Hall effect, where counterpropagating spin-polarized edge states coexist at opposite boundaries of an insulating bulk~\cite{PhysRevLett.95.226801,chang2009optical,doi:10.1126/science.1148047,doi:10.1143/JPSJ.77.031007}. These helical edge modes are protected by time-reversal symmetry (\(\mathcal{T}\)) and are immune to nonmagnetic backscattering, leading to dissipationless spin currents and quantized edge conductance of \(2e^2/h\)~\cite{PhysRevLett.97.036808,RevModPhys.82.3045,RevModPhys.83.1057}.
                       
                       Theoretical proposals of the QSH effect in graphene~\cite{PhysRevLett.95.226801} demonstrated the essential role of spin–orbit coupling (SOC) in generating a nontrivial topological gap, though the small SOC in graphene limited experimental realization. This challenge was overcome by Bernevig, Hughes, and Zhang in HgTe/CdTe quantum wells, where a thickness-tuned band inversion produced a measurable QSH phase~\cite{Bernevig2006}. The experimental observation of the QSH effect in these wells~\cite{doi:10.1126/science.1148047} inaugurated the era of 2D topological insulators (TIs), motivating extensive efforts to identify new material platforms with large band gaps, chemical stability, and compatibility with devices~\cite{PhysRevLett.100.236601,Ando}.
                       
                       In the ensuing decade, the QSH state has been realized or proposed in a wide variety of 2D systems, including InAs/GaSb quantum wells~\cite{PhysRevLett.107.136603,PhysRevLett.114.096802}, silicene and germanene~\cite{PhysRevLett.107.076802,Ezawa_2012}, bismuth bilayers~\cite{PhysRevLett.97.236805,Reis}, stanene~\cite{PhysRevLett.111.136804}, and particularly TMDs in their distorted 1T$^{\prime}$ phase~\cite{Xia,Tang2017,Fei2017}. The topological character of these materials is typically characterized through the \(\mathbb{Z}_2\) invariant~\cite{Xia} or the spin Chern number~\cite{Tang2017}, both of which relate to the BC distribution and band inversion near the Fermi level.
                       
                       The bulk–edge correspondence implies that a nontrivial topological invariant guarantees the presence of gapless helical edge states traversing the bulk gap~\cite{RevModPhys.82.3045,RevModPhys.83.1057}. These edge modes yield quantized spin Hall conductance and are robust against time reversal preserving perturbations, contrasting with the quantum anomalous Hall effect (QAHE), where \(\mathcal{T}\) symmetry is broken by magnetism~\cite{Rui,Wang_2015,Rachel_2018}. The interplay between SOC, crystal symmetry, and inversion breaking perturbations determines the onset of QSH behavior, and recent work has emphasized the tunability of these properties via strain, electric fields, or light–matter coupling~\cite{PhysRevB.84.195430,PhysRevLett.111.146802,PhysRevB.79.081406,Wang2013}.
                       
                       In parallel, the discovery of atomically thin van der Waals materials has provided an ideal playground for realizing 2D topological phases. Among them, TMDs with formula MX$_2$ (M = Mo, W; X = S, Se, Te) have attracted enormous attention due to their rich polymorphism and strong SOC~\cite{Chhowalla2013,Wang2012,Manzeli2017}. The semiconducting 2H phase of MoS$_2$ is the most widely studied, exhibiting a direct band gap of about 1.8~eV in the monolayer limit and valley-selective circular dichroism~\cite{PhysRevLett.105.136805,PhysRevLett.108.196802}. However, metastable metallic 1T and semimetallic 1T$^{\prime}$ phases of MoS$_2$ exhibit drastically different properties, including Peierls distortions, charge density wave instabilities, and topologically nontrivial band orderings~\cite{Heising1999}.
                       
                       The distorted 1T$^{\prime}$ phase, characterized by a zigzag chain of Mo atoms, has been theoretically predicted to host a robust QSH state with an energy gap on the order of 0.1–0.3~eV, suitable for room temperature operation~\cite{Xia}. First-principles calculations revealed that strong SOC and lattice distortion drive band inversion between Mo-\(d\) and S-\(p\) orbitals near the \(\Gamma\) point, resulting in a nontrivial \(\mathbb{Z}_2\) topology~\cite{Xia,lu2017janus}. Experimental signatures of this topological insulating phase have been reported in monolayer WTe$_2$~\cite{Fei2017,Wu2018}, and related observations in Mo-based TMDs suggest a shared underlying mechanism~\cite{Chen2019,Katsuragawa2020}. In 1T$^{\prime}$-MoS$_2$, the magnitude of the SOC and the presence of inversion asymmetry further enrich its spin and valley physics, making it a compelling platform for exploring QSH and valley Hall effects~\cite{Xu2023}.
                       
                       The ability to manipulate the topological character of 1T$^{\prime}$-MoS$_2$ using external fields or optical driving opens exciting prospects for nonequilibrium topological phenomena. Floquet engineering has emerged as a powerful tool to dynamically induce and tune topological states~\cite{Kitagawa,Mikami2016}. When a 1T$^{\prime}$-MoS$_2$ monolayer is subjected to a circularly polarized light (CPL), the effective Hamiltonian acquires light-induced mass terms that modify the BC and band topology, leading to Floquet–Bloch bands with controllable Chern numbers~\cite{Dehghani,Claassen2016,Hubener2017}. Such light-driven transitions between trivial, QSH, and quantum anomalous Hall phases have been theoretically demonstrated in several 2D materials~\cite{Perez2014,Foa2014,Usaj2014}, making 1T$^{\prime}$-MoS$_2$ an ideal candidate for tunable topological spintronics.
                       
                           \begin{figure*}
                       	\centering
                       	\begin{subfigure}[b]{0.49\textwidth}
                       		\centering
                       		\includegraphics[width=\linewidth]{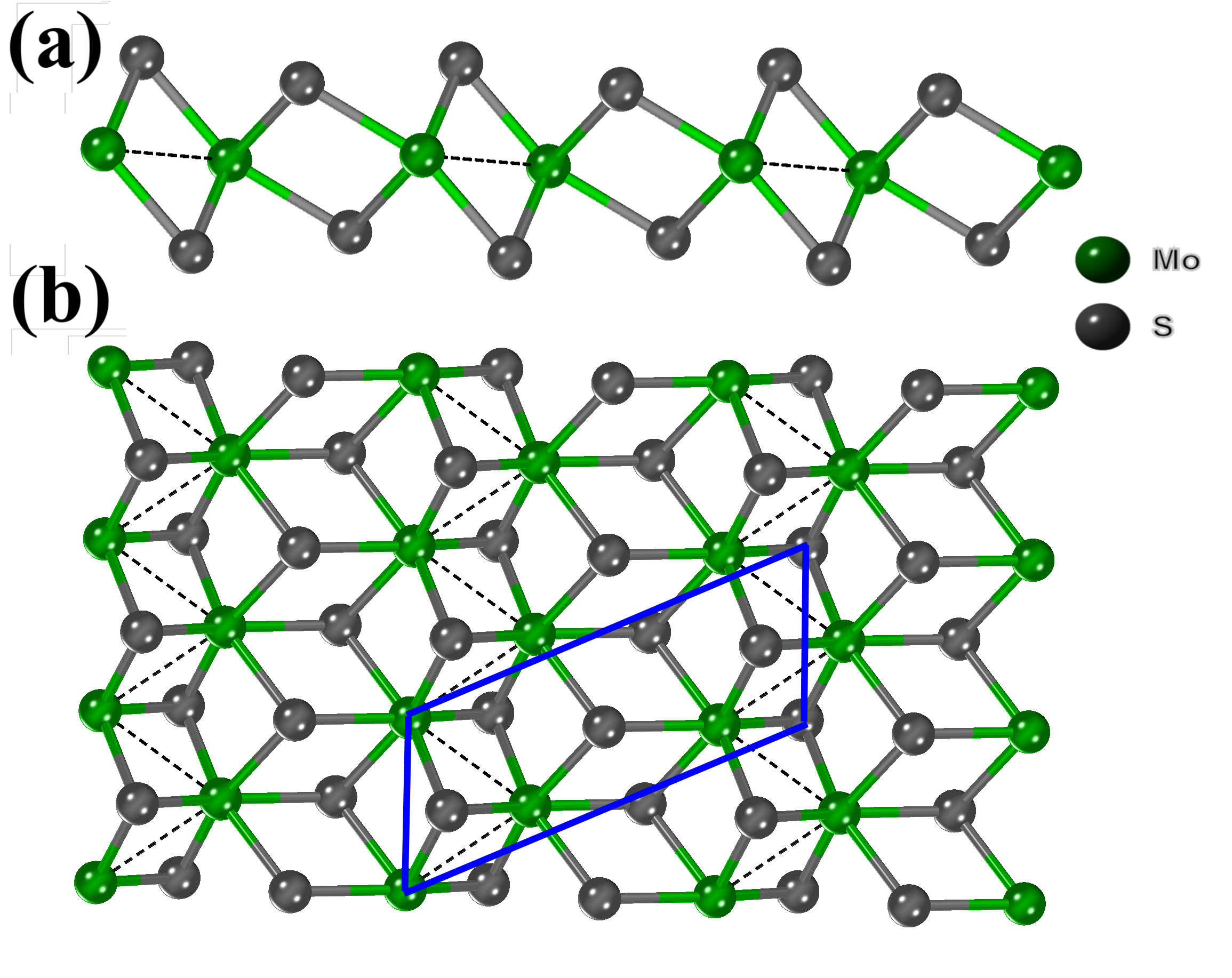 }
                       		
                       		\label{fig:1a}
                       	\end{subfigure}
                       	\hfill
                       	\begin{subfigure}[b]{0.50\textwidth}
                       		\centering
                       		\includegraphics[width=\linewidth]{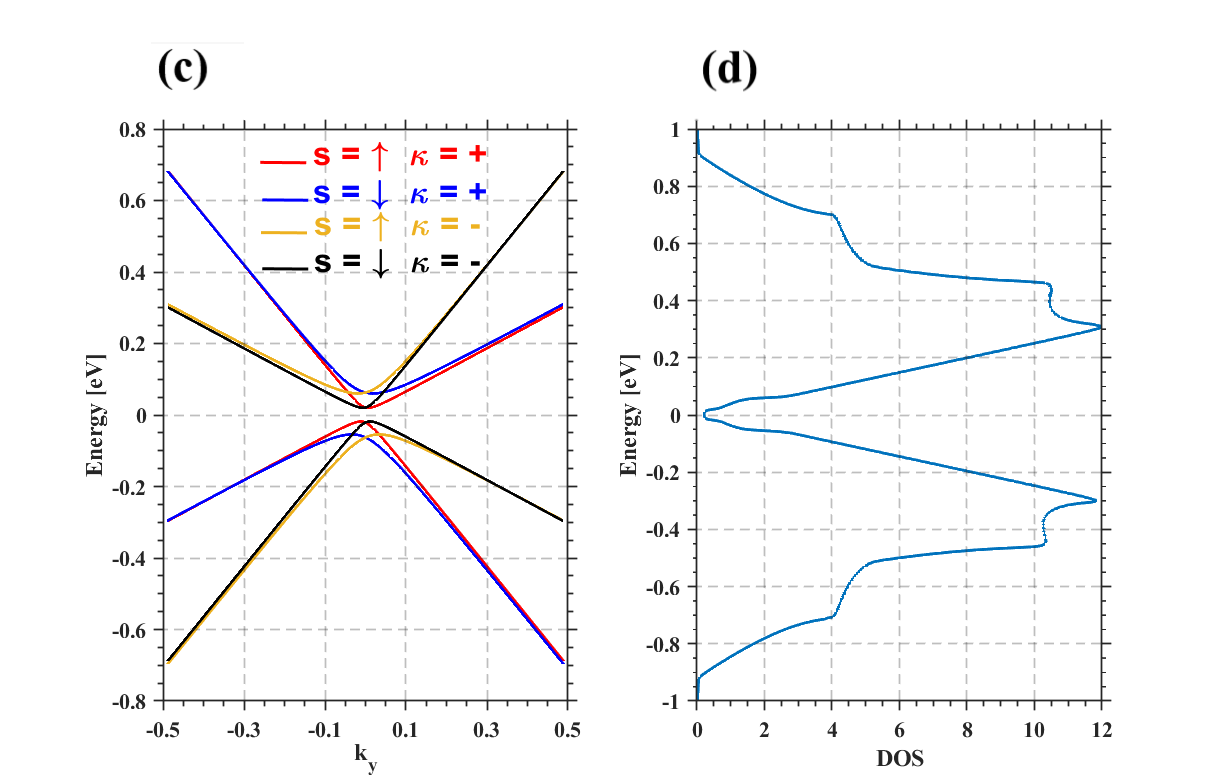}
                       		
                       		\label{fig:1b}
                       	\end{subfigure}
                       	
                       	\caption{(a) Side view and (b) top view of the crystal structure of monolayer 1T$^{\prime}$-MoS$_2$. Molybdenum (Mo) and sulfur (S) atoms are represented by green and dark-gray spheres, respectively. The dashed lines highlight the Mo–Mo dimerization pattern, and the blue parallelogram marks the primitive unit cell used in our calculations.(c) Spin- and valley-resolved band structure of monolayer 1T$^{\prime}$-MoS$_2$ near the Fermi level, showing the effects of SOC and valley asymmetry. 
                       		The red and blue curves correspond to the spin-up and spin-down states at the $\kappa = +$ valley, while the yellow and black curves represent the same at the $\kappa = -$ valley. 
                       		A small gap opening at $k_y = 0$ reflects SOC-induced band splitting. 
                       		(d) Corresponding DOS illustrating a small minimum near the Fermi energy and pronounced van Hove singularities on both sides, consistent with the anisotropic band structure of the 1T$'$ phase.}
                       	\label{fig:1}
                       \end{figure*}

                      Previous studies have explored  and Floquet engineering in monolayer $1T'$-MoS$_2$, demonstrating that periodic optical driving can substantially modify its electronic band structure, anisotropy, and topological properties. In particular, first-principles calculations have established light-induced transitions between quantum spin Hall, quantum anomalous Hall, and topologically trivial phases in $1T'$ transition-metal dichalcogenides \cite{Kong_2022}, while subsequent investigations have focused on the polarization-dependent modification of the tilted Dirac bands and their energy gaps in $1T'$-MoS$_2$ \cite{Iurov2022}. Building on these developments, the present work provides a systematic interplay between a circularly polarized light and a perpendicular electric field within a  low-energy $k\cdot p$ Hamiltonian. In particular, we derive the analytical Floquet-induced mass corrections and establish explicit expressions for the Chern number, BC and the critical parameters at which individual spin-valley sectors undergo a topological phase transition. By systematically evaluating the BC and spin-valley-resolved Chern numbers, we construct the nonequilibrium topological phase diagram and demonstrate how the competition between electric-field and light-induced time-reversal-symmetry breaking controls the transitions among the QSHl, BI, S-QHI, and P-QHI phases. 
                       
     
     
 
     
    



     
     
 
     
    



\section{Theory\label{sec:2}}
Figures~\ref{fig:1}(a) and \ref{fig:1}(b) depict the side and top views of monolayer 1T$'$-MoS$_2$, respectively. This two-dimensional material comprises two atomic species combined of molybdenum (Mo), shown in green, and sulfur (S), shown in dark-gray and the primitive unit cell, highlighted by a green parallelogram.

The low-energy $\boldsymbol{k} \cdot \boldsymbol{p}$ Hamiltonian of monolayer 1T$'$-MoS$_2$ in the $x$–$y$ plane, subject to a perpendicular electric field $E_z$ and gate voltage $V$, has been previously established in Refs.~\cite{Nobahari2025,Xia, Das2020, PhysRevB.107.035301}. The total Hamiltonian is expressed as the sum of the intrinsic Hamiltonian $H_{k \cdot p}$, the electric field contribution $H_{E_z}$, and the gate potential $V$:
\begin{equation}
	H = H_{k \cdot p} + H_{E_z} + V.\label{eq:1}
\end{equation}

The unperturbed Hamiltonian $H_{k \cdot p}$ is given by:
\begin{equation}
	H_{k \cdot p} =
	\begin{pmatrix}
		E_p & 0 & -i \hbar \nu_1 q_x & \hbar \nu_2 q_y \\
		0 & E_p & \hbar \nu_2 q_y & -i \hbar \nu_1 q_x \\
		i \hbar \nu_1 q_x & \hbar \nu_2 q_y & E_d & 0 \\
		\hbar \nu_2 q_y & i \hbar \nu_1 q_x & 0 & E_d
	\end{pmatrix},\label{eq:2}
\end{equation}
and the contribution from the electric field is:
\begin{equation}
	H_{E_z} = \alpha \Delta_{\mathrm{SO}}
	\begin{pmatrix}
		0 & 0 & 1 & 0 \\
		0 & 0 & 0 & 1 \\
		1 & 0 & 0 & 0 \\
		0 & 1 & 0 & 0
	\end{pmatrix},\label{eq:3}
\end{equation}
where $V$ is a scalar gate potential, and $\Delta_{\mathrm{SO}} = 41.9~\mathrm{meV}$ denotes the intrinsic SOC gap at the Dirac points $(0, \kappa \Lambda)$. The dimensionless parameter $\alpha = |E_z / E_c|$, with $E_c = 1.42~\mathrm{V/nm}$ being the critical electric field required to induce a topological phase transition.

The on-site energies for the $p$ and $d$ orbitals are defined as:
\begin{align}
	E_p &= \delta_p + \frac{\hbar^2 q_x^2}{2 m^p_x} + \frac{\hbar^2 q_y^2}{2 m^p_y},\label{eq:4} \\
	E_d &= \delta_d + \frac{\hbar^2 q_x^2}{2 m^d_x} + \frac{\hbar^2 q_y^2}{2 m^d_y},\label{eq:5}
\end{align}
where $\delta_p = 0.46~\mathrm{eV}$, $\delta_d = -0.20~\mathrm{eV}$, $m^p_x = -0.50 m_0$, $m^p_y = -0.16 m_0$, $m^d_x = 2.48 m_0$, $m^d_y = 0.37 m_0$, and $m_0$ is the free electron mass. The Fermi velocities along the $x$ and $y$ directions are $\nu_1 = 3.87 \times 10^5~\mathrm{m/s}$ and $\nu_2 = 0.46 \times 10^5~\mathrm{m/s}$, respectively.

After applying a unitary transformation, the effective Hamiltonian in the spin and valley basis becomes:

	\begin{equation} 
	\begin{split}
		H_{\kappa,s}(k)&=\hbar k_x\nu_1\sigma_y-\hbar k_y(s\nu_2\sigma_x+\kappa\nu_{-}\sigma_0 \\ 
		&+\kappa\nu_{+}\sigma_z)+\Delta_{so}(\alpha-s\kappa)\sigma_x+V\sigma_{0}\label{eq:6}
	\end{split}
\end{equation}

where $s = \pm 1$ denotes spin-up and spin-down states, and $\kappa = \pm 1$ labels the $K$ and $K'$ valleys, respectively. The tilting velocities $\nu_{\pm}$ are given by:
\begin{align}
	\nu_{-} &= \frac{\hbar q_0}{2} \left( -\frac{1}{m^p_y} - \frac{1}{m^d_y} \right) = 2.84 \times 10^5~\mathrm{m/s},\label{eq:7} \\
	\nu_{+} &= \frac{\hbar q_0}{2} \left( -\frac{1}{m^p_y} + \frac{1}{m^d_y} \right) = 7.18 \times 10^5~\mathrm{m/s}.\label{eq:8}
\end{align}

This effective Hamiltonian captures the anisotropic band structure and tunable topological properties of the 1T$'$-MoS$_2$ monolayer under external electric field.

It is convinient to write Hamiltonian as $H=d(k).\sigma+d_0\sigma_0$. After diagonalization of the Hamiltonian we have

\begin{align}
	\varepsilon_{\kappa,s}^{\pm}(k_x,k_y)&=d_0(k_y) \nonumber\\
	&\pm\sqrt{[d_x(k_y)]^2+[d_y(k_x)]^2+[d_z(k_y)]^2}\label{eq:9}
\end{align}

and corresponding eigenstates are
\begin{equation}
	\psi^{\pm}_{\kappa,s}(k_x,k_y)=N_{\pm}
\begin{pmatrix}
	\dfrac{d_x(k_y)-i\,d_y(k_x)}{\pm|\mathbf{d}|-d_z(k_y)}\\[2ex]
	1
\end{pmatrix},
\end{equation}
where
\begin{equation*}
N_{\pm}=\left[\,1+\frac{\left|d_x(k_y)-i\,d_y(k_x)\right|^{2}}{\left(\pm|\mathbf{d}|-d_z(k_y)\right)^{2}}\right]^{-1/2},\label{eq:10}
\end{equation*}

where $d_x^0(k_y)=\Delta_{so}(\alpha-s\kappa)-\hbar k_ys\nu_2$, $d_y^0(k_x)=\hbar k_x\nu_1$, $d_z^0(k_y)=-\hbar k_y\kappa\nu_{+}$, $d_0(k_y)=V-\hbar k_y\kappa\nu_-$, and the mass vectors are $M_x^0=d_x^0(0)=\Delta_{so}(\alpha-s\kappa)$, $M_y^0=d_y^0(0)=0$, $M_z^0=d_z^0(0)=0$.

\subsection{Coupling to a periodic drive}
We consider spatially-uniform circularly polarized light in the dipole approximation. The vector potential is
\begin{equation}
	\mathbf{A}(t) = A_0\big(\cos(\omega t),\,\eta\sin(\omega t)\big),\label{eq:11}
	\end{equation}
	\begin{equation*}
	\eta=\pm1 \ \text{(right/left helicity)},
\end{equation*}
and minimal coupling is implemented via $\mathbf{k}\mapsto\mathbf{k}+\frac{e}{\hbar}\mathbf{A}(t)$.

Retaining terms linear in $\mathbf{A}$ the time-dependent Hamiltonian becomes
\begin{equation}
	H_{\kappa s}(t) = H_{\kappa s}(k) + e\,\nu_1 A_x(t)\,\sigma_y - e\,A_y(t)\,\Pi_{\kappa s},\label{eq:12}
\end{equation}
where we defined
\begin{equation}
	\Pi_{\kappa s} \equiv s\,\nu_2\,\sigma_x + \kappa\,\nu_{-}\,\sigma_0 + \kappa\,\nu_{+}\,\sigma_z .\label{eq:13}
\end{equation}
We consider the off-resonant high-frequency regime, in which the photon energy exceeds the relevant interband energy scales of the effective model. This condition suppresses real photon-induced transitions and allows the periodically driven system to be described by an effective static Hamiltonian obtained through the van Vleck expansion.
For a $T=2\pi/\omega$-periodic Hamiltonian the Schrödinger equation admits Floquet solutions $\Psi_\alpha(t)=e^{-i\epsilon_\alpha t/\hbar}\Phi_\alpha(t)$ with $\Phi_\alpha(t+T)=\Phi_\alpha(t)$. In the off-resonant regime, The Floquet Hamiltonian can be written as
\begin{equation}
	H_F = H^{(0)} + \frac{1}{\hbar\omega}\,[H_{-1},H_{+1}] + \mathcal{O}(\omega^{-2}),\label{eq:14}
\end{equation}

\begin{equation}
	H_{(\pm1)} \equiv \frac{1}{T}\int_0^T dt\, H(t)e^{\pm i\omega t},\label{eq:15}
\end{equation}

where $H_{\pm1}$ are the $\pm\omega$ Fourier components of $H(t)$.

Using $A_x(t)=\frac{A_0}{2}\big(e^{i\omega t}+e^{-i\omega t}\big)$ and $A_y(t)=\frac{\eta A_0}{2i}\big(e^{i\omega t}-e^{-i\omega t}\big)$

the Fourier components from Eq.~\eqref{eq:15} are
\begin{align}
	H_{+1} &= \frac{eA_0}{2}\Big(\nu_1\sigma_y + \frac{\eta}{i}\Pi_{\kappa s}\Big)\label{eq:6}, \\
	H_{-1} &= \frac{eA_0}{2}\Big(\nu_1\sigma_y - \frac{\eta}{i}\Pi_{\kappa s}\Big)\label{eq:17}.
\end{align}

A direct commutator evaluation (using $[\sigma_y,\sigma_x]=-2i\sigma_z$ and $[\sigma_y,\sigma_z]=2i\sigma_x$) gives
\begin{equation}
	[H_{-1},H_{+1}] = e^2\nu_1\,\eta\,A_0^2\left(\kappa\,\nu_{+}\,\sigma_x-s\,\nu_2\,\sigma_z  \right)\label{eq:18}.
\end{equation}

Substituting Eq.~\eqref{eq:18} into Eq.~\eqref{eq:14} yields the effective Floquet Hamiltonian to leading order in $1/\omega$:
\begin{equation}
	\delta H_F^{(\kappa s)} = \frac{e^2\nu_1 A_0^2}{\hbar \omega}\,\eta \left(\kappa \nu_{+} \sigma_x- s\nu_2 \sigma_z  \right)\label{eq:19} .
\end{equation}

and total Hamiltonian is
\begin{equation}
\begin{split}
	H_F^{(\kappa s)}(\mathbf{k}) \approx\; &
	\hbar k_x \,\nu_1 \,\sigma_y
	- \hbar k_y\!\left(s\,\nu_2\,\sigma_x + \kappa\,\nu_{-}\,\sigma_0  +\kappa\,\nu_{+}\,\sigma_z\right)\\
	& + \Big[\Delta_{\mathrm{so}}(\alpha - s\kappa) + \xi\,\nu_+\,\eta\,\kappa\Big]\sigma_x\\
	&-\xi\,\nu_2\eta\,s\,\sigma_z
	+ V\sigma_0.
\end{split} \label{eq:20}
\end{equation}
where we defined $\xi \equiv\frac{e^2\nu_1}{\hbar\omega}\,A_0^2$. 

\subsection{Mass vector, spectrum and light-shifted criticality}
It is convenient to write $H_F = \mathbf{d^{\prime}}(\mathbf{k})\cdot\boldsymbol{\sigma} + d_0(\mathbf{k})\sigma_0$ with components
\begin{align}
	d^{\prime}_x(\mathbf{k}) &= -\hbar k_y\,s\nu_2 + \Delta_{\mathrm{so}}(\alpha - s\kappa) + \xi\,\nu_+\eta\,\kappa\label{eq:21},\\
	d^{\prime}_y(\mathbf{k}) &= \hbar k_x\,\nu_1\label{eq:22},\\
	d_z^{\prime}(\mathbf{k}) &= -\hbar k_y\,\kappa\nu_+ - \xi\,\nu_2\,\eta\,s\label{eq:23},\\
	d_0(\mathbf{k}) &= V - \hbar k_y\,\kappa\nu_-\label{eq:24}.
\end{align}
The quasienergies are $\epsilon^{\pm}(\mathbf{k}) = d_0(\mathbf{k}) \pm \sqrt{d_x'^2+d_y'^2+d_z'^2}$.

At $\mathbf{k}=0$ the drive produces two Floquet shifted effective masses contributions $M_x \equiv d'_x(0) = \Delta_{\mathrm{so}}(\alpha - s\kappa) + \xi\,\nu_+\eta\,\kappa$, $M_y\equiv d'_y(0)=0$, $M_z \equiv d'_z(0) = -\xi\,\nu_2\,\eta\,s$.

Therefore the quasieigenvalues at $\mathbf{k}=0$ are $ \epsilon^{\pm}(0)=V\pm\sqrt{M_x^2+M_z^2}$.

A direct band touching requires all three components of $\mathbf d'$ to vanish. From $d_y'=\hbar k_x\nu_1=0$, we obtain $k_x=0$, while $d_x'=d_z'=0$ require
\begin{equation}
	M_x-\hbar k_y s\nu_2=0,\qquad M_z-\hbar k_y\kappa\nu_+=0\label{eq:25}.
\end{equation}
Eliminating $k_y$ gives the gap-closing condition
\begin{equation}
	\Gamma\equiv\kappa\nu_+M_x-s\nu_2M_z=0\label{eq:26}.
\end{equation}
Substituting the masses yields
\begin{equation}
	\Gamma=\kappa\nu_+\Delta_{\mathrm{so}}(\alpha-s\kappa)+\eta\xi(\nu_+^2+\nu_2^2)\label{eq:27},
\end{equation}
so that, for a fixed spin, valley, helicity, and electric field,
\begin{equation}
	\xi_c=-\frac{\kappa\nu_+\Delta_{\mathrm{so}}(\alpha-s\kappa)}{\eta(\nu_+^2+\nu_2^2)}\label{eq:28}.
\end{equation}
Equivalently, the critical shifted electric-field parameter satisfies
\begin{equation}
	\alpha_c^{(\kappa s)}=s\kappa-\frac{\eta\xi(\nu_+^2+\nu_2^2)}{\kappa\nu_+\Delta_{\mathrm{so}}}\label{eq:29}.
\end{equation}

\subsection{BC of the Floquet Hamiltonian}

For a generic two-band Hamiltonian of the form
\begin{equation}
	H(\mathbf k) = d_0(\mathbf k)\,\sigma_0 + \mathbf d'(\mathbf k)\cdot\boldsymbol\sigma , \label{eq:30}
\end{equation}
with $\mathbf d'(\mathbf k)=(d'_x,d'_y,d'_z)$, the BC of the upper ($+$) and lower ($-$) bands is given by
\begin{equation}
	\Omega_\pm^z(\mathbf k) \;=\; \mp\,\frac{1}{2}\,
	\frac{\mathbf d' \cdot \big( \partial_{k_x}\mathbf d' \times \partial_{k_y}\mathbf d' \big)}
	{|\mathbf d'|^3},
	\qquad |\mathbf d'| = \sqrt{d_x'^2+d_y'^2+d_z'^2} \label{eq:31}.
\end{equation}
The scalar term $d_0(\mathbf k)$ only shifts the energy and does not contribute. Evaluating the derivatives,
\begin{equation}
	\partial_{k_x}\mathbf d'=(0,\hbar\nu_1,0), 
	\qquad
	\partial_{k_y}\mathbf d'=(-\hbar s\nu_2,\,0,\,-\hbar\kappa\nu_+) \label{eq:32},
\end{equation}
one finds
\begin{equation}
	\mathbf d' \cdot \left(\partial_{k_x}\mathbf d' \times \partial_{k_y}\mathbf d' \right)
	= -\hbar^2\nu_1\left(\kappa\nu_+ M_x - s\nu_2 M_z\right) \label{eq:33},
\end{equation}
which is independent of $\mathbf k$.

\medskip

The final BC of the lower band is therefore
\begin{equation}
	{\;
		\Omega_-^{z}(\mathbf k) \;=\;
		-\frac{\hbar^2\nu_1}{2}\,
		\frac{\kappa\nu_+ M_x - s\nu_2 M_z}
		{|d'|^3}
		\;} \label{eq:34}
\end{equation}
where 
\begin{equation}
|d'|^3=\big[(M_x-\hbar k_y s\nu_2)^2 + (\hbar k_x\nu_1)^2 + (M_z-\hbar k_y\kappa\nu_+)^2 \big]^{3/2} \label{eq:35}
\end{equation}
while the upper band has the opposite sign,
\begin{equation}
	\Omega_+^{z}(\mathbf k) = -\,\Omega_-^{z}(\mathbf k) \label{eq:36}.
\end{equation}

\medskip

Note that the tilt velocity $\nu_-$ and scalar potential $V$ appear only in $d_0$ and therefore
do not affect the BC. The sign and magnitude of the curvature are controlled by the
effective Floquet masses $M_x$ and $M_z$, which depend explicitly on the light helicity $\eta$
and amplitude $A_0$.

\subsection{Hall conductivity}

We begin from the BC of the valence Floquet band for a given valley $\kappa=\pm1$ and spin $s=\pm1$,
\begin{equation}
	\Omega_-^{z}(\mathbf k)
	= -\frac{\hbar^2\nu_1}{2}\;
	\frac{B M_x - A M_z}
	{|d'|^3} \label{eq:37},
\end{equation}
where, for compactness, we defined
\[
A \equiv s\,\nu_2,\qquad B \equiv \kappa\,\nu_{+},
\]

The Hall conductivity for sector $(\kappa,s)$ is
\begin{equation}
	\sigma_{xy}^{(\kappa s)} \;=\; \frac{e^2}{\hbar}\int_{\mathbb R^2}\!\frac{d^2k}{(2\pi)^2}\;
	\Omega^{z}(\mathbf k)
	\;=\; \frac{e^2}{h}\,C_{\kappa s} \label{eq:38},
\end{equation}
with the Chern number
\begin{equation}
C_{\kappa s} \;=\; \frac{1}{2\pi}\int_{\mathbb R^2} d^2k\; \Omega^{z}(\mathbf k) \label{eq:39}.
\end{equation}

The Chern number formula can be written as (see Appendix \ref{sec:5} )
   \begin{equation}
	{\;
		C_{\kappa s} \;=\; -\frac{1}{2}\,\operatorname{sgn}\!\big(\kappa\nu_+\,M_x - s\nu_2\,M_z\big)\;.} \label{eq:40}
\end{equation}  

Each valley–spin Dirac sector contributes a half-integer, $C_{\kappa,s} = \pm1/2$. Integer values of the Chern numbers in Figs. 3–4 and Tables I–II arise after summing over the sectors. Accordingly $C_{spin}$ and $C_{valley}$ can take half-integer values.

Let \(C_{\kappa s}\) denote the Chern number of the occupied band manifold in valley \(\kappa=\pm\) and spin channel \(s=\uparrow,\downarrow\).
The total Chern number is then
\begin{equation}
	C_{\mathrm{tot}}(\xi)
	= \sum_{\kappa=\pm}\sum_{s=\uparrow,\downarrow} C_{\kappa s}
	= \sum_{s} C_{s}(\xi)
	= \sum_{\kappa} C_{\kappa}(\xi)\label{eq:41},
\end{equation}
where
\(C_{s}(\xi)=\sum_{\kappa}C_{\kappa s}\) and
\(C_{\kappa}(\xi)=\sum_{s}C_{\kappa s}\)
are the spin- and valley-resolved Chern numbers, respectively.
We further define
\begin{equation}
	C_{\mathrm{valley}}=\frac{C_{+}-C_{-}}{2}, \qquad\label{eq:42}
 C_{\mathrm{spin}}=\frac{C_{\uparrow}-C_{\downarrow}}{2}.
\end{equation}

\begin{figure}
	\centering
	\includegraphics[width=\linewidth]{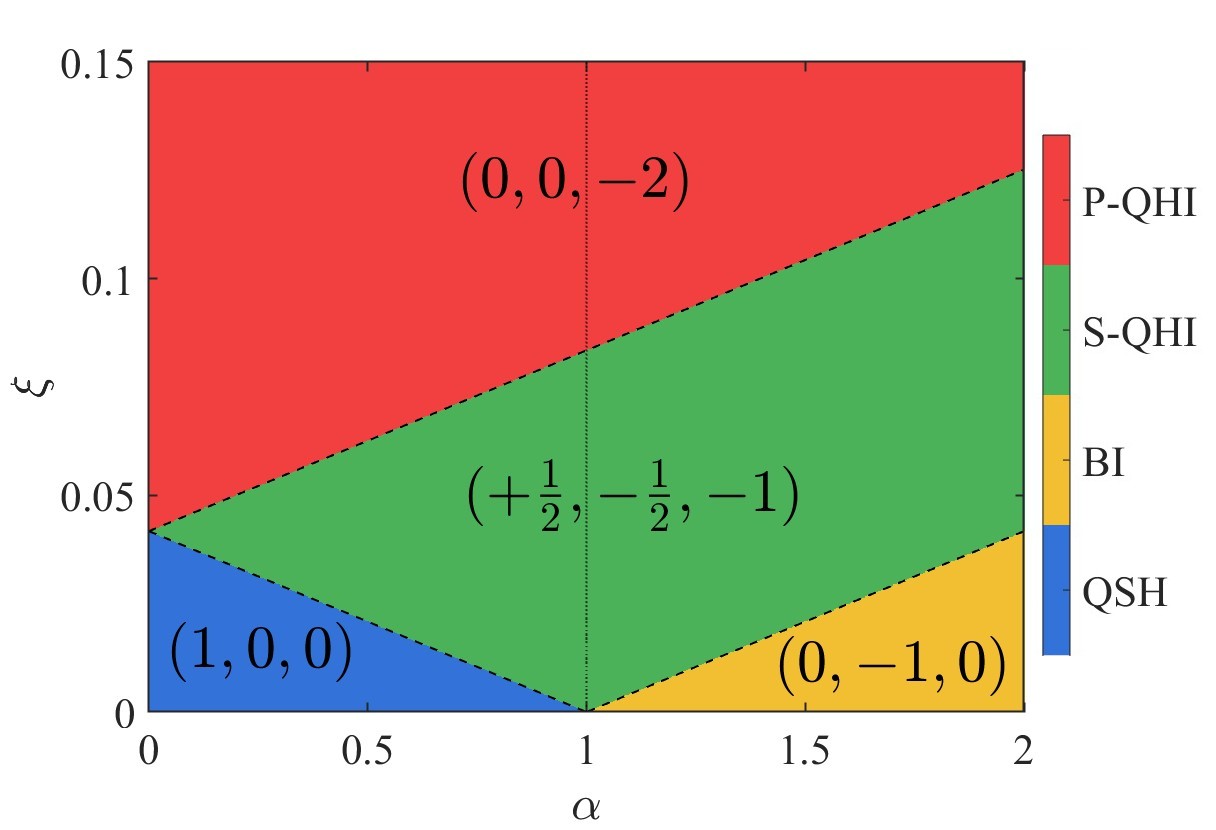 }
	\caption{Topological phase diagram of Floquet-driven monolayer 1T$^{\prime}$-MoS$_2$ in the ($\alpha$, $\xi$) plane, for circularly polarized light with $\eta = +1$. The colors mark the QSH (blue), BI (yellow), S-QHI (green) and P-QHI (red) phases. Dashed lines are the analytic gap-closing boundaries, $\Gamma_{\kappa,s} = 0$. The dotted vertical line at $\alpha = 1$ marks the field-induced QSH–BI transition of the undriven system. The values indicated within each region correspond to $(C_{\mathrm{spin}},C_{\mathrm{valley}},C_{\mathrm tot})$.}\label{fig:2}
\end{figure}

\begin{figure*}
	\centering
	\begin{subfigure}[b]{0.3\textwidth}
		\centering
		\includegraphics[width=\linewidth]{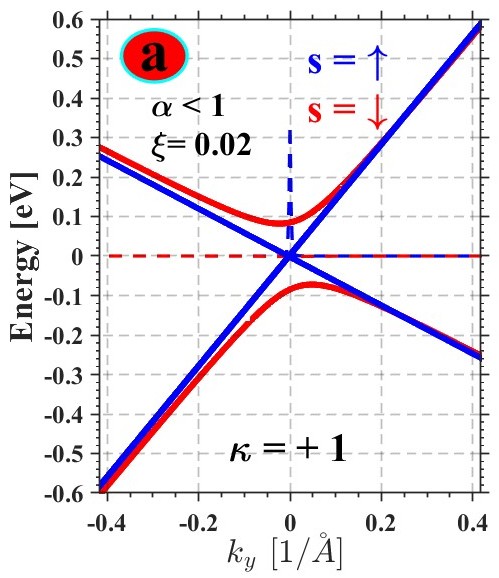 }
		
		\label{fig:3a}
	\end{subfigure}
	\hfill
	\begin{subfigure}[b]{0.3\textwidth}
		\centering
		\includegraphics[width=\linewidth]{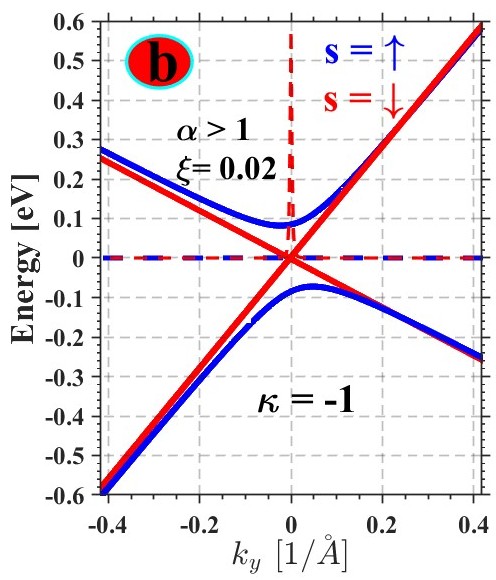}
		
		\label{fig:3b}
	\end{subfigure}
	\hfill
	\begin{subfigure}[b]{0.3\textwidth}
		\centering
		\includegraphics[width=\linewidth]{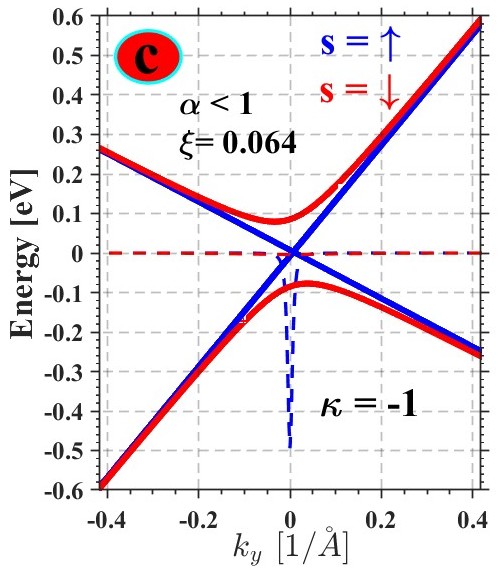}
		
		\label{fig:3c}
	\end{subfigure}
	
	\vspace{10pt} 
	
	\begin{subfigure}[b]{0.3\textwidth}
		\centering
		\includegraphics[width=\linewidth]{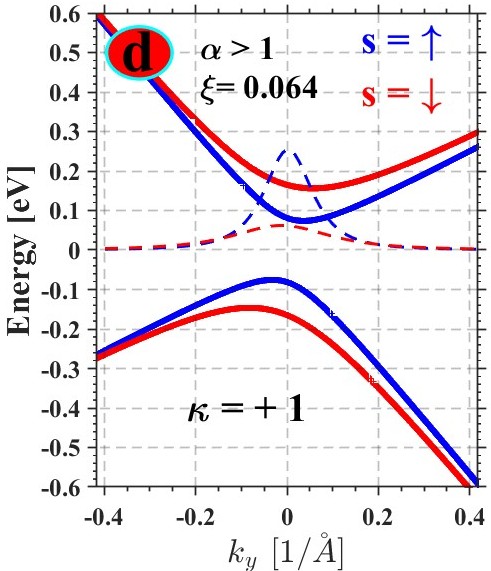}
		
		\label{fig:3d}
	\end{subfigure}
	\hfill
	\begin{subfigure}[b]{0.3\textwidth}
		\centering
		\includegraphics[width=\linewidth]{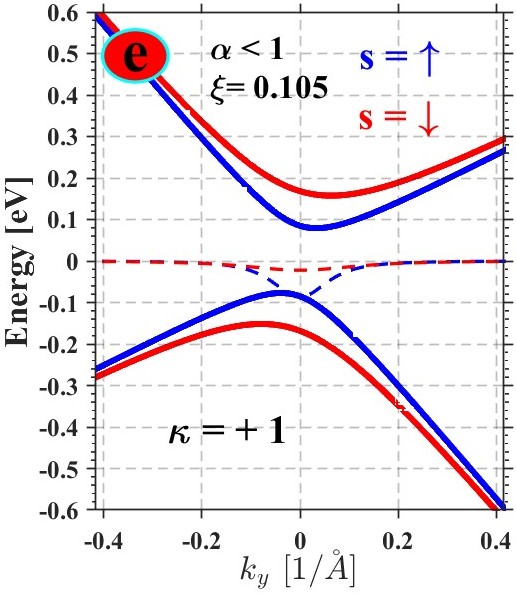}
		
		\label{fig:3e}
	\end{subfigure}
	\hfill
	\begin{subfigure}[b]{0.3\textwidth}
		\centering
		\includegraphics[width=\linewidth]{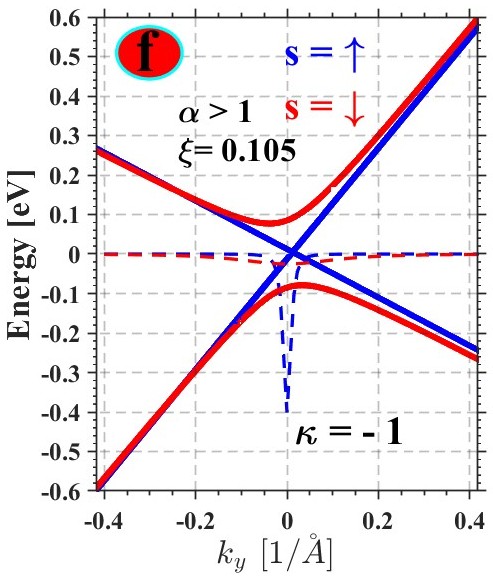}
		
		\label{fig:3f}
	\end{subfigure}
	\caption{Floquet-engineered band structures and corresponding BC of monolayer 1T$'$--MoS$_2$ for different spin and valley indices under various light--matter coupling strengths ($\xi$) and electric field ($\alpha$).
		Panels (a--f) display the quasienergy spectra (solid lines) and BC distributions (dashed lines) for spin-up (blue) and spin-down (red) states at valleys $\kappa = \pm 1$. 
		Panels (a,b) correspond to $\xi \simeq 0.02$ and (c,d) to $\xi \simeq 0.064$, while (e,f) show results for $\xi \simeq 0.105$.}
	\label{fig:3}
\end{figure*}

\begin{table}[h]
	\centering
	\caption{Spin- and valley-resolved Chern numbers versus Floquet driving strength for $\alpha=1.5$.}
	\begin{tabular}{c|cccccccc}
		\hline\hline
		Floquet range & $C_{\uparrow}$  &$C_{\downarrow}$  &$C_{+}$ & $C_{-}$ & $C_{\mathrm{tot}}$ & $ C_{\mathrm{spin}}$ & $C_{\mathrm{valley}}$ & Phase  \\
		\hline
		$\xi<0.02$ & $0$ & $0$ & $-1$ & $+1$ & $0$ & $0$ & $-1$ & BI\\
		$0.02<\xi<0.105$ & $0$ & $-1$ & $-1$ & $0$ & $-1$ & $+1/2$ & $-1/2$ & S-QHI \\
		$\xi>0.105$ & $-1$ & $-1$ & $-1$ & $-1$ & $-2$ & $0$ & $0$ & P-QHI \\
		\hline\hline\label{tabel:1}
	\end{tabular}
\end{table}

\begin{table}[h]
	\centering
	\caption{Spin- and valley-resolved Chern numbers versus Floquet driving strength for $\alpha=0.5$.}
	\begin{tabular}{c|cccccccc}
		\hline\hline
		Floquet range & $C_{\uparrow}$  &$C_{\downarrow}$  &$C_{+}$ & $C_{-}$ & $C_{\mathrm{tot}}$ & $ C_{\mathrm{spin}}$ & $C_{\mathrm{valley}}$ & Phase \\
		\hline
		$\xi<0.02$ & $+1$ & $-1$ & $0$ & $0$ & $0$ & $+1$ & $0$ & QSH\\
		$0.02<\xi<0.064$ & $0$ & $-1$ & $-1$ & $0$ & $-1$ & $+1/2$ & $-1/2$ & S-QHI \\
		$\xi>0.064$ & $-1$ & $-1$ & $-1$ & $-1$ & $-2$ & $0$ & $0$ & P-QHI \\
		\hline\hline
	\end{tabular}
\end{table}

\begin{figure}[]
	\centering
	\begin{tabular}{cc}
		\includegraphics[width=0.5\linewidth]{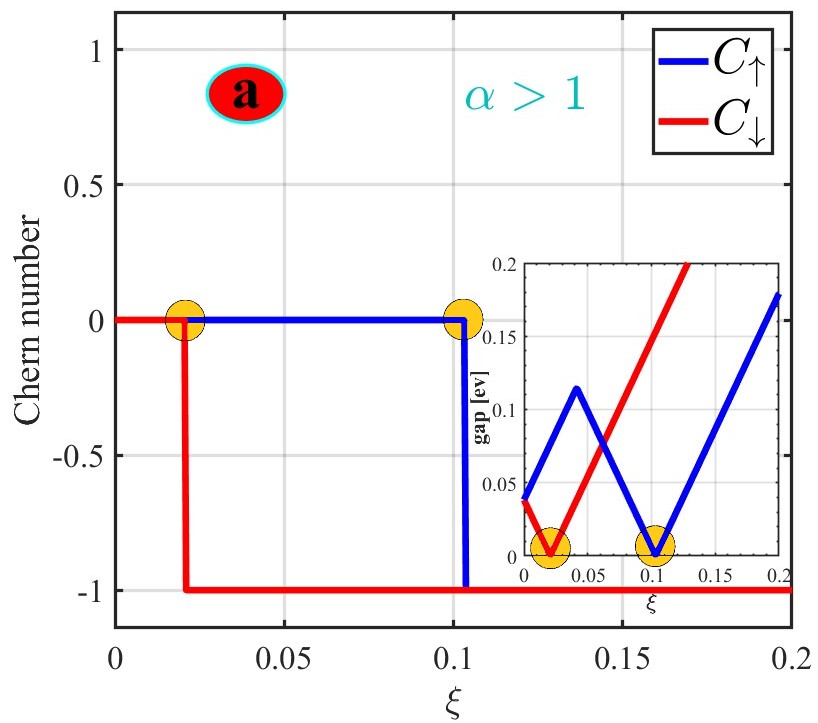}&
		\includegraphics[width=0.5\linewidth]{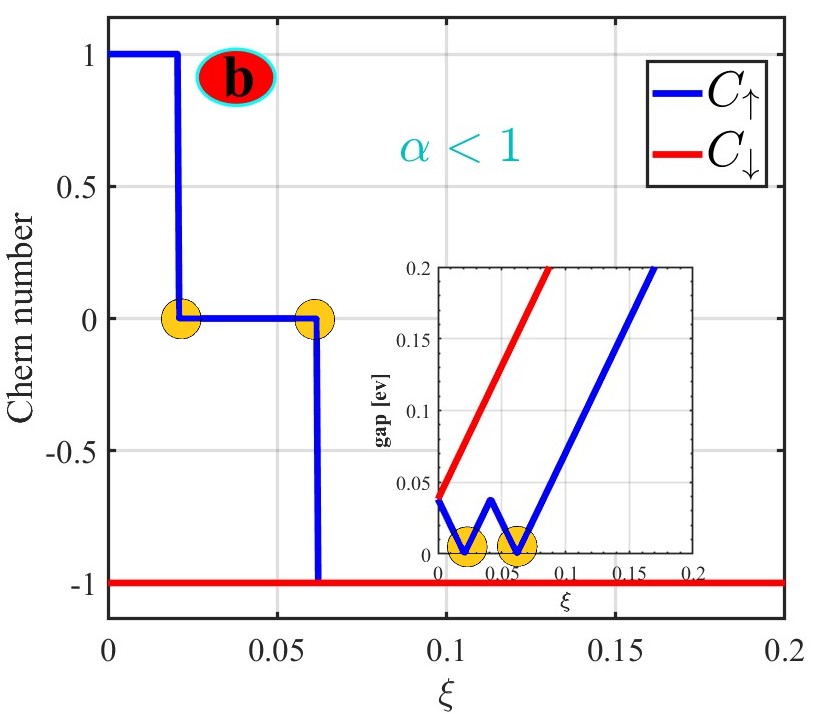}\\[2\tabcolsep]
		\includegraphics[width=0.5\linewidth]{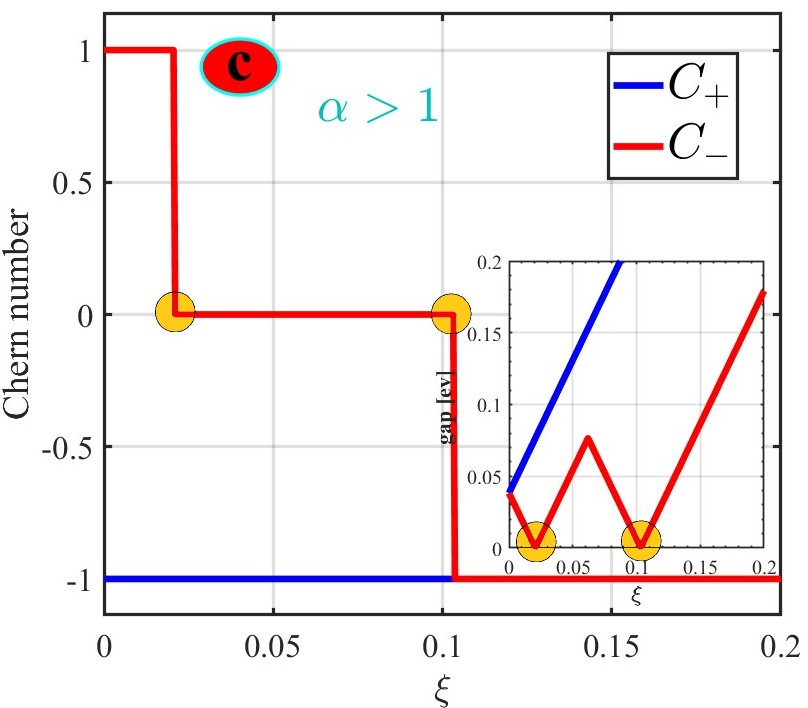}&
		\includegraphics[width=0.5\linewidth]{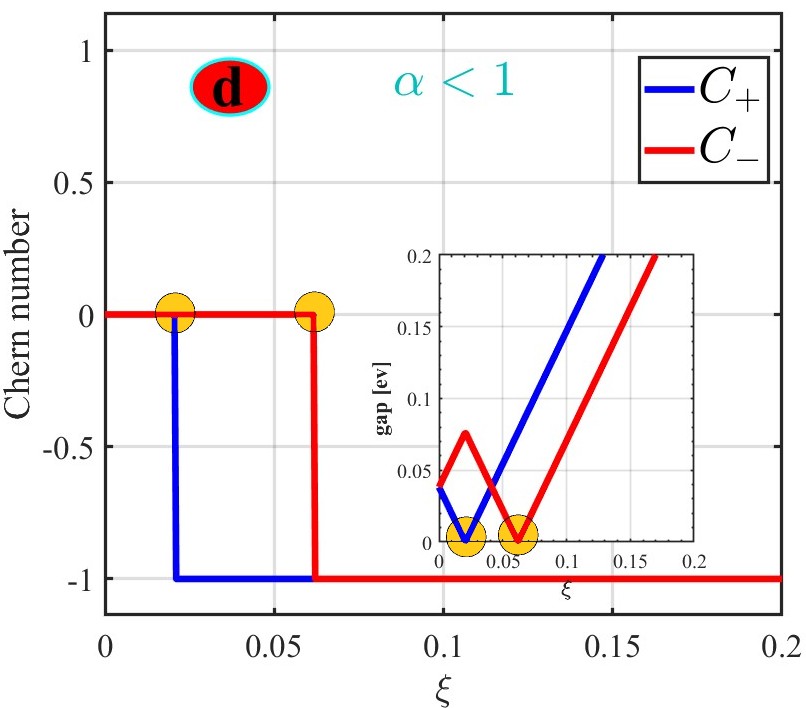}
	\end{tabular}
	\caption{Spin and valley Chern numbers of monolayer 1T$'$--MoS$_2$ as a function of the Floquet coupling strength ($\xi$) for different electric field amounts.
		Panels (a,b) show the spin Chern numbers $C_{\uparrow}$ and $C_{\downarrow}$, while panels (c,d) present the valley Chern numbers $C_{+}$ and $C_{-}$ for $\alpha > 1$ and $\alpha < 1$, respectively. 
		The insets illustrate the variation of the light-induced band gap with $\xi$ for each spin and valley configuration. 
	}\label{fig:4}
\end{figure}

\section{Results and Discussion\label{sec:3}} 

Figures~\ref{fig:1}(c) and \ref{fig:1}(d) display the low-energy band dispersion and the corresponding density of states (DOS) for monolayer 1T$'$--MoS$_2$ near the Fermi level, revealing the essential electronic features that underpin its topological behavior. Here velocities are normalized in respect to $\nu_+$ and $\nu_1$ so we have $\nu_+=1$, $\nu_-=0.3955$, $\nu_1=1$ and $\nu_2=0.064$.
In Fig.~\ref{fig:1}(c), the band structure is resolved according to spin ($S = \uparrow, \downarrow$) and valley ($\kappa = \pm$) indices, highlighting the spin--valley coupling inherent to the 1T$'$ phase. 
The red and blue curves correspond to spin-up and spin-down states at the $\kappa = +$ valley, respectively, whereas the yellow and black curves represent the spin-up and spin-down branches at the $\kappa = -$ valley.

At $k_y = 0$, a small band gap opens at the crossing point, evidencing the presence of SOC, which lifts the degeneracy and separates the spin-resolved branches. 
The asymmetry between the spin-up and spin-down dispersions reflects the broken inversion symmetry of the system combined with strong SOC originating from the Mo $d$ orbitals. 
The linear crossing of the conduction and valence bands near the Fermi level indicates Dirac-like behavior modified by SOC, which transforms the massless Dirac fermions into massive ones and results in a topological insulating phase depending on the sign of the band inversion parameter.

Figure~\ref{fig:1}(d) presents the corresponding DOS, which is consistent with the features observed in the band structure. 
Near the Fermi energy ($E = 0$), the DOS exhibits a minimum corresponding to the small band gap. 
The sharp peaks on both sides of the Fermi level arise from van Hove singularities associated with the nearly flat portions of the bands at higher energies. 

Figure~\ref{fig:2} summarizes the topological phase diagram of the driven system as function of the vertical electric field parameter $\alpha$ and Floquet coupling $\xi$. In absence of the laser perturbation ($\xi=0$), around $\alpha=1$, the gap closes and reopens and the system undergoes a topological phase transition from QSH insulator with $(C_{\uparrow}, C_{\downarrow})= (+1, -1)$ to the BI phase with $(C_+, C_-)= (-1, +1)$ and total Chern number $C_{tot}=0$. 

Turning on the circularly polarized drive breaks time-reversal symmetry and produces net Chern numbers. Both equilibrium phases are stable only for small $\xi$. They occupy two triangular regions bounded by $\xi = \xi_0(1- \alpha)$ for $\alpha < 1$ and $\xi = \xi_0 (\alpha -1)$ for $\alpha > 1$, where $\xi_0=\nu_+\Delta_{so}/(\nu_+^2+\nu_2^2)\simeq 0.042$. Crossing these lines closes the gap in a single spin–valley sector and drives the system into the S-QHI, with $C_{tot} = -1$. This phase is the green region, and it is bounded above by the line $\xi = \xi_0(1+\alpha)$. Beyond this line, a second sector inverts and the system enters the  P-QHI, with $(C_{\uparrow}, C_{\downarrow}) = (-1, -1)$ and $C_{tot} = -2$.

For $\eta=+1$, only three of the four spin–valley sectors can invert, $(\kappa, s) = (+, \uparrow)$ at $\xi = \xi_0(1-\alpha)$, $(-, \downarrow)$ at $\xi = \xi_0(\alpha-1)$, and $(-, \uparrow)$ at $\xi = \xi_0(1+\alpha)$. The $(+, \downarrow)$ sector has $s - \kappa\alpha = - (1+\alpha) <0$ and is never inverted. The phase sequence is therefore QSH $\rightarrow$ S-QHI $\rightarrow$ P-QHI for $\alpha<1$ and BI $\rightarrow$ S-QHI $\rightarrow$ P-QHI for $\alpha>1$.

The field-induced transition at $\alpha=1$ exists only at $\xi=0$ and is hidden by the drive for any $\xi>0$, so the QSH and BI states both evolve into the same S-QHI phase. Additionally the width of the S-QHI window depends on $\alpha$. For $\alpha>1$ it is constant, equal to $2\xi_0 \simeq 2\nu_+\Delta_{so}/(\nu_+^2+\nu_2^2)$ and for $\alpha < 1$ it is $2\alpha\xi_0$ and closes at $\alpha=0$. At that point the $(+, \uparrow)$ and $(-, \uparrow)$ sectors become degenerate and invert together, so $C_{\uparrow}$ changes from +1 to -1 in a single transition, and the system passes from QSH directly to P-QHI at $\xi=\xi_0$. For $\alpha<1$, increasing $\alpha$ widens the S-QHI window . For $\alpha>1$, the S-QHI window has the constant width of $2\alpha\xi_0$, but its lower edge $\xi_0(\alpha-1)$ moves up with $\alpha$, so a larger field makes the S-QHI phase harder to reach from the BI side.

Reversing the helicity ($\eta =-1$) flips the sign of the induced masses and exchanges the roles of the spin–valley sectors. The phase boundaries are mirrored, and the Chern numbers of the driven phases change sign ($C_{tot} =+1$ and +2). Thus, the handedness of the light controls the sign of the induced Hall response.

Figure~\ref{fig:3} (a-f) summarizes the Floquet quasienergy dispersions and the corresponding BC distributions of monolayer 1T$'$–MoS$_2$ for representative values of the light–matter coupling $\xi$, the electric field $\alpha$, and valley index $\kappa$ for different spin configurations. The magnitude of the BC are renormalized to fit in the figures. Figure~\ref{fig:3} (a) show the weak-driving regime, $\xi=0.02$, which lies close to the first critical coupling for both values of $\alpha$. The spin-up bands in the $\kappa=+1$ valley touch linearly at $k_y \approx0$, while the spin-down bands remain gapped about 0.15 eV. This touching marks the first Floquet-induced band inversion, in the $(\kappa =+, s = \uparrow)$ sector at $\xi_c =\xi_0(1-\alpha)$, and it is accompanied by a sharp and strongly localized BC peak in the spin-up channel. The spin-down curvature is negligible on the same scale. Across this point $C_{\uparrow}$ changes from +1 to 0 and the system phase changes from QSH phase to the S-QHI phase. For$\alpha>1$, (see \ref{fig:3} (b)) , there is almost same condition but the gap closes in the $\kappa =-1$ valley, in the spin-down sector at $\xi_c=\xi_0(\alpha-1)$, with a corresponding sharp spin-down BC. Here $C_{\downarrow}$ changes from 0 to -1 and the system goes from the BI phase to the S-QHI phase. Panels (a) and (b) therefore show that the first inversion occurs at a similar value of $\xi$ in both regimes, but the electric-field parameter $\alpha$ determines which spin–valley sector inverts.
 Figures \ref{fig:3}(c) and \ref{fig:3}(d) show the intermediate coupling $\xi=0.064$. For $\alpha<1$, the $\kappa=-1$ valley shows a linear crossing in the spin-up sector, together with a sharply localized BC peak of opposite sign to that in panel (a). This corresponds to the second critical coupling, $\xi_c=\xi_0(1+\alpha)$, where $C_{\uparrow}$ changes from 0 to -1 and the system enters the P-QHI phase with $(C_{\uparrow}, C_{\downarrow})=(-1,-1)$. For $\alpha>1$, panel \ref{fig:3} (d), the $\kappa=+1$ valley has no critical point in this range of $\xi$, and both spin branches remain gapped.
 Figures~\ref{fig:3} (e) and (f) show the strongest driving, $\xi=0.105$. For $\alpha<1$, panel \ref{fig:3}(e), the $\kappa=+1$ valley is fully gapped, with gaps of about 0.16 eV for spin-up and 0.31 eV for spin-down. Both spin branches have passed their critical couplings, so the gap is now set by the light-induced mass rather than by the intrinsic SOC. The BC is weak and spread over a wide range of $k_y$, which reflects the large gap. This is the robust bulk gap that supports the quantized Hall plateaus of the P-QHI phase. For $\alpha>1$, panel \ref{fig:3} (f), the $\kappa=-1$ valley again shows a band crossing at $k_y\approx0$ with a sharply localized BC peak. This is the second inversion for $\alpha>1$, in the $(-,\uparrow)$ sector at $\xi_c=\xi_0(1+\alpha)$, and it drives the transition from S-QHI to P-QHI. Panels (c) and (f) thus show the same sector closing in both regimes, at a critical coupling that rises linearly with $\alpha$.

Figure \ref{fig:4}(a) shows the spin Chern numbers for $\alpha>1$ and inset represents the gap versus the $\xi$. In the absence of light, both channels are trivial, $C_{\uparrow}=C_{\downarrow}=0$. At $\xi=0.02$, the spin-down gap closes and reopens, and $C_{\downarrow}$ changes from 0 to -1, while $C_{\uparrow}$ remains zero. This inversion takes place in the $\kappa=-$ valley and marks the transition from the BI phase to the S-QHI, with a total Chern number $C_{tot}=-1$. A second transition occurs at $\xi=0.105$, where the spin-up gap closes and $C_{\uparrow}$ also drops from 0 to -1. Beyond this point, $(C_{\uparrow}, C_{\downarrow})= (-1, -1)$ and $C_{tot}=-2$, which defines the P-QHI. The insets confirm that the two transitions occur in different spin channels, with the spin-down gap closing first and the spin-up gap closing later.
Figure \ref{fig:4}(b) shows the same quantities for $\alpha<1$. Here the undriven system is a QSH insulator with $(C_{\uparrow}, C_{\downarrow})= (+1 -1)$ and $C_{tot}=0$. At $\xi=0.02$, the spin-up gap closes in the $\kappa=+$ valley, and $C_{\uparrow}$ changes from +1 to 0. The system then enters the S-QHI phase, with $(C_{\uparrow}, C_{\downarrow})= (0, -1)$ and $C_{tot}=-1$. At $\xi=0.064$, a second closing in the $\kappa=-$ valley takes $C_{\uparrow}$ from 0 to -1, and the system reaches the P-QHI phase. The spin-down channel remains at $C_{\downarrow}=-1$ throughout, and its gap increases monotonically with $\xi$ without closing. All the topological changes for $\alpha<1$ are therefore confined to the spin-up channel. The sequence QSH $\rightarrow$ S-QHI $\rightarrow$ P-QHI is thus driven by two successive inversions of the same spin species, in opposite valleys.
Figure \ref{fig:4}(c) shows the valley Chern numbers for $\alpha>1$. The undriven state is a time-reversal-symmetric, valley-contrasting phase with $C_+=-1$ and $C_-=-1$, so that $C_{tot}=0$. The $\kappa=+$ valley stays at $C_+=-1$ over the entire range, and its gap grows steadily with $\xi$. In contrast, the $\kappa=-$ valley undergoes two transitions. At $\xi=0.02$, its Chern number changes from +1 to 0, leaving BI state in which only the $\kappa=+$ valley carries a chiral contribution. At $\xi=0.105$, $C_-$ change again, from 0 to -1, so that both valleys carry the same Chern number and the system enters the P-QHI phase. The inset of the $\kappa=-$ valley shows two gap closings which reflects the topological phase transitions.

Figure \ref{fig:4}(d) shows the valley Chern numbers for $\alpha<1$. Both valleys are trivial at weak driving, $C_+=C_-=0$. This is consistent with the QSH phase, where each valley hosts one spin-up and one spin-down sector with opposite Chern numbers that cancel within the valley. At $\xi=0.02$, the gap of the $\kappa=+$ valley closes and $C_+$ changes from 0 to -1. The $\kappa=-$ valley remains trivial up to $\xi=0.064$, where its gap closes and $C_-$ also becomes -1. The intermediate region is therefore a valley-polarized phase, and the final state is the P-QHI phase with $(C_+, C_-)=(-1, -1)$.

\section{Conclusions\label{sec:4}}
The results of this work delineate how circular driving restructures the spin--valley landscape of 1T$^\prime$--MoS$_2$ and organizes its nonequilibrium topological phases. The Floquet-renormalized spectrum exhibits a sequence of drive-controlled gap closings, each accompanied by a redistribution of BC among the spin and valley sectors. These critical points define boundaries separating quantum spin Hall, BI, S-QHI, and P-QHI regimes. The spin- and valley-resolved Chern numbers reveal that the response of individual sectors can be tuned independently through the Floquet coupling strength and the electric-field. This tunability originates from distinct light-induced mass renormalizations acting differently on the two valleys and spin channels. Overall, the results identify 1T$^\prime$--MoS$_2$ as a material whose topological transport properties can be reconfigured continuously and reversibly by high-frequency driving, highlighting its potential for light-programmable topological functionalities.

 \section{Appendix} \label{sec:5}                
  
   Substitute \eqref{eq:37} into the Chern integral:
   \begin{equation}
   	C_{\kappa s}
   	=- \frac{1}{2\pi}\,\frac{\hbar^2\nu_1}{2}\,(B M_x - A M_z)
   	\int d^2k\;
   	\frac{1}{|d|^3} \label{eq:43}.
   \end{equation}
   
   Introduce the linear variables
   \[
   u \equiv \hbar \nu_1 k_x, \qquad v \equiv \hbar k_y.
   \]
   Then
   \[
   d^2k = dk_x\,dk_y = \frac{du\,dv}{\hbar^2\nu_1},
   \]
   and the denominator becomes
   \[
   D(u,v) = (M_x - A v)^2 + u^2 + (M_z - B v)^2.
   \]
   Thus
   \begin{equation}
   	C_{\kappa s}
   	=- \frac{1}{2\pi}\,\frac{\hbar^2\nu_1}{2}\,(B M_x - A M_z)
   	\int \frac{du\,dv}{\hbar^2\nu_1}\;\frac{1}{[D(u,v)]^{3/2}}\label{eq:44}.
   \end{equation}
   Cancel the $\hbar^2\nu_1$ prefactors:
   \begin{equation}
   	C_{\kappa s}
   	= -\frac{1}{4\pi}\,(B M_x - A M_z)
   	\int du\,dv\;\frac{1}{[D(u,v)]^{3/2}}\label{eq:45}.
   \end{equation}
   
   Write the quadratic form in $v$:
   \[
   D(u,v) = u^2 + (A^2+B^2)\,v^2 - 2v(A M_x + B M_z) + M_x^2 + M_z^2.
   \]
   Complete the square by defining
   \[
   \tilde w \equiv \sqrt{A^2+B^2}\; v - \frac{A M_x + B M_z}{\sqrt{A^2+B^2}}.
   \]
   Then
   \[
   D(u,v) = u^2 + \tilde w^2 + M_x^2 + M_z^2 - \frac{(A M_x + B M_z)^2}{A^2+B^2}.
   \]
   Define the effective mass
   \[
   M_{\rm eff}^2 \equiv M_x^2 + M_z^2 - \frac{(A M_x + B M_z)^2}{A^2+B^2}.
   \]
   A short algebraic rearrangement shows
   \[
   M_{\rm eff}^2
   = \frac{(B M_x - A M_z)^2}{A^2+B^2}.
   \]
   (Indeed this identity is the numerator/denominator algebra behind the simplification of the integral.)
   
   The Jacobian for $(u,v)\mapsto(u,\tilde w)$ is $|\partial(u,v)/\partial(u,\tilde w)| = 1/\sqrt{A^2+B^2}$ since $\tilde w=\sqrt{A^2+B^2}\,v - \mathrm{const}$. Therefore
   \begin{equation}
   	\int du\,dv\;\frac{1}{[D(u,v)]^{3/2}}= \frac{1}{\sqrt{A^2+B^2}}\int du\,d\tilde w\;\frac{1}{\big[u^2+\tilde w^2 + M_{\rm eff}^2\big]^{3/2}}\label{eq:46}.
   \end{equation}

   Let $r^2 = u^2 + \tilde w^2$, $d u\, d\tilde w = r\,dr\,d\theta$. The angular integral gives a factor $2\pi$:
   \[
   \int du\,d\tilde w\;\frac{1}{(r^2+M_{\rm eff}^2)^{3/2}}
   = 2\pi \int_0^{\infty} \frac{r\,dr}{(r^2+M_{\rm eff}^2)^{3/2}}.
   \]
   Evaluate the radial integral (standard):
   \[
   \int_0^\infty \frac{r\,dr}{(r^2+a^2)^{3/2}} = \frac{1}{a}.
   \]
   Hence
   \[
   \int du\,dv\;\frac{1}{[D(u,v)]^{3/2}}
   = \frac{1}{\sqrt{A^2+B^2}}\; 2\pi\; \frac{1}{M_{\rm eff}}.
   \]
   
   Using $M_{\rm eff} = \dfrac{|B M_x - A M_z|}{\sqrt{A^2+B^2}}$ we obtain the closed form
   \[
   \int du\,dv\;\frac{1}{[D(u,v)]^{3/2}} = 2\pi\;\frac{1}{|B M_x - A M_z|}.
   \]

   Substitute back into the expression for $C_{\kappa s}$:
   \begin{equation}
   	\begin{split}
   		C_{\kappa s}&
   		= -\frac{1}{4\pi}\,(B M_x - A M_z)\; \Big[2\pi\;\frac{1}{|B M_x - A M_z|}\Big]
   		\\
   		&= -\frac{1}{2}\,\operatorname{sgn}\!\big(B M_x - A M_z\big).
   	\end{split}\label{eq:47}
   \end{equation}
   
   Recalling $A=s\nu_2$ and $B=\kappa\nu_+$, this is
   
   \begin{equation}
   	{\;
   		C_{\kappa s} \;=\; -\frac{1}{2}\,\operatorname{sgn}\!\big(\kappa\nu_+\,M_x - s\nu_2\,M_z\big)\;.}\label{eq:48}
   \end{equation}      
            
\section*{Conflict of interests}
            The authors have no conflicts to disclose.

\bibliography{ref}
\end{document}